\documentclass[11pt,twoside]{article}

\usepackage{asp2006}
\usepackage{epsf}
\usepackage{lscape}

\begin{document}
\title{New Light Curves and Spectra of the close PG1159 Binary System
  \mbox{SDSS J212531.92$-$010745.9}}   
\author{Sonja Schuh,$^1$ Thorsten Nagel,$^2$ Iris Traulsen,$^1$ and Benjamin Beeck$^1$}   
\affil{$^1$Institut f\"ur Astrophysik,  Georg-August-Universit\"at G\"ottingen,
  Friedrich-Hund-Platz 1, 37077 G\"ottingen, Germany}    
\affil{$^2$Institut f\"ur Astronomie und Astrophysik, Eberhard-Karls-Universit\"at T\"ubingen,
  Sand 1, 72076 T\"ubingen, Germany}    
\begin{abstract} 
  Methods to measure masses of PG\,1159 stars in order to test
  evolutionary scenarios are currently based on spectroscopic masses
  or asteroseismological mass
  determinations. \mbox{SDSS~J212531.92$-$010745.9}, a recently
  discovered PG\,1159 star in a close binary system, may finally allow
  the first dynamical mass determination, and has so far been analysed
  on the basis of one SDSS spectrum and photometric monitoring.
  In order to be able to phase radial velocity measurements of the
  system \mbox{SDSS~J212531.92$-$010745.9}, we have followed
  up the photometric monitoring of this system. New white light time
  series of the brightness variation of
  \mbox{SDSS~J212531.92$-$010745.9} with the T\"ubingen 80\,cm and
  G\"ottingen 50\,cm telescopes extend the monitoring into a second
  season (2006). We give the improved ephemeris for the orbital motion
  of the system, based on a sine fit which now results in a period of
  6.95573(5)\,h.
  In 2007, we have obtained a series of phase-resolved medium-resolution
  spectra with the TWIN spectrograph at the 3.5\,m telescope at Calar
  Alto, which will allow us to derive the radial velocity curves for
  both components of the system, and to perform spectral analyses of
  the irradiating and irradiated components at different phases. We
  briefly describe the newly obtained spectra.
  The light curve and radial velocities combined will allow us
  to carry out a mass determination.
\end{abstract}
%
%
%
The difficulties in measuring masses for PG\,1159 stars, using
spectroscopic and asteroseismic methods, are discussed in Werner et
al.\ (these proceedings), and references therein. 
Their conclusion is that current errors in mass determinations are typically
larger than 15\,\%, resulting in even larger uncertainties for the
progenitor masses. 
\par
While the role of binary stellar evolution is highly
relevant in the discussion of the possible progenitors for many
types of hydrogen-deficient stars, binarity does not seem to be a
particularly common occurence among PG\,1159 stars (and is not needed
to explain their origin). In fact, \mbox{SDSS~J212531.92$-$010745.9}
(discovered by \citealt*{2006A&A...448L..25N}) remains the only
securely confirmed PG\,1159 star with a close companion.
\par
The significance of this system however derives less from its
contribution to binary statistics, and much more from the fact that it
contains the first PG\,1159 star for which a dynamical mass
determination will be possible \citep*{2006astro.ph.10324S}. This will
add an important third method of mass determination to the
spectroscopic and asteroseimic approaches.
\par
The light curve of \mbox{SDSS~J212531.92$-$010745.9} shows a 6.96\,h
variation that \citet{2006A&A...448L..25N} interpret as the orbital
period of the system, with the periodic brightening caused by a
reflection effect. Based on white light photometric measurements
obtained at the T\"ubingen 80\,cm and G\"ottingen 50\,cm telescopes,
spanning two observing seasons, \cite{2008astro.ph.00000S}
report the improved ephemeris for the times of maximum light to be
\[
\begin{array}{lr@{}lcr@{}l}
  &  {\rm HJD} = 2454055\fd 213&4(4)  & + &    0\fd 28982&2(2) \cdot E. \\
\end{array}
\]
The associated new amplitude determination, obtained from a sine fit,
results in 0.299(33) relative intensity change. The phased light
curve variation profile, with its shape deviating from strictly sinusoidal,
reinforces the interpretation of the variation as a reflection 
effect.
\par
Spectroscopic observations of \mbox{SDSS~J212531.92$-$010745.9} were
obtained on 2007/08/20 and 08/21 at the 3.5\,m telescope at Calar Alto,
equipped with the TWIN spectrograph and gratings T08 for the blue
and T04 for the red channel. At an exposure time of 1800\,s, a
total of 13 usable spectra were obtained within two nights, resulting
in a good phase coverage.
A first look at the spectra clearly confirms that the photometric variation
indeed corresponds to the orbital period, as the same period is seen
in the preliminary radial velocity curves. Furthermore, the strength
of the hydrogen emission lines, evident in the snapshot provided by
the SDSS spectrum, and attributed to the irradiated side of the
companion \citep{2006A&A...448L..25N}, varies with phase as expected.
\par
We will be confronted with two problems when interpreting the value of
the dynamic mass in a broader context and comparing it with results
for spectroscopic and asteroseismic masses: The spectroscopic
parameters for\linebreak \mbox{SDSS~J212531.92$-$010745.9}, in particular
$\log{g}$, will still suffer from the same uncertainties (due to
uncertainties in the line broadening theory) as those of other
PG\,1159 stars; and furthermore, since no pulsations have so far been
detected in this object, a  cross-calibration of asteroseismic mass
determinations will not be immediately possible.
\acknowledgements 
We would like to thank
Elke Reiff,
Derek Homeier,
Heike Schwager,
Daniel-Jens Kusterer,
Ronny Lutz,
Sylvia Brandert,
Sebastian Wende,
Jens Adamczak,
Agnes Hoffmann,
Ralf Kotulla,
Simon H\"ugel\-meyer,
Markus Hundertmark,
and Johannes Fleig for supporting the photometric observations in 2006;
and Klaus Reinsch for encouragement and technical support.
The spectroscopy is based on observations collected in service
mode by Javier Aceituno and Ulrich Thiele at the Centro Astron\'{o}mico Hispano
Alem\'{a}n (CAHA) at Calar Alto, operated jointly by the Max-Planck 
Institut f\"ur Astronomie and the Instituto de Astrof\'{\i}sica de Andaluc\'{\i}a
(CSIC). 
We gratefully acknowledge the initialization of this project by Boris
G\"ansicke and Matthias Schreiber who first directed our attention to
this special object.
I.~Traulsen acknowledges a travel grant by the DLR under project
number 50\,OR\,0501.
B.~Beeck thanks the conference organizers for generous support. 
%
%
%

%

\begin{thebibliography}{}
\bibitem[Nagel et al.(2006)]{2006A&A...448L..25N} Nagel, T., Schuh, S., 
Kusterer, D.-J., Stahn, T., H{\"u}gelmeyer, S.~D., Dreizler, S., 
G{\"a}nsicke, B.~T., \& Schreiber, M.~R.\ 2006, \aap, 448, L25 
\bibitem[Schuh \& Nagel(2007)]{2006astro.ph.10324S} Schuh, S., \& Nagel, 
T.\ 2007, in ASP Conf.\ Ser.\ Vol.\ 372, 15$^{\textrm{th}}$ European Workshop on
White Dwarfs, eds.\ R.~Napiwotzki \& M.~R.~Burleigh (San Francisco: ASP),  491
\bibitem[Schuh et al.(2007)]{2008astro.ph.00000S} Schuh, S., Traulsen,
  I., Nagel, T., Reiff, E., Homeier, D., Schwager, H., Kusterer,
  D.-J., Lutz, R., Schreiber, M.~R.\ 2008, AN, accepted
\end{thebibliography}
\end{document}